\documentclass[aps,prl,showpacs,twocolumn]{revtex4-1}
\usepackage{graphicx,amssymb,amsmath,epsfig,color}

\begin{document}
\title{Finite-Temperature Fluid-Insulator Transition of Strongly Interacting 1D Disordered Bosons}

\author{V.P. Michal$^{1}$, I.L. Aleiner$^{2}$, B.L. Altshuler$^{2,5}$, and G.V. Shlyapnikov$^{1,3,4,5}$}
\affiliation{\mbox{$^{1}$ Laboratoire de Physique Th{\'e}orique et Mod{\`e}les Statistiques,  Universit{\' e} Paris Sud, CNRS, 91405 Orsay, France}\\
\mbox{$^{2}$Physics Department, Columbia University, 538 West 120th Street, New York, New York 10027, USA}\\
\mbox{$^{3}$Van der Waals-Zeeman Institute, University of Amsterdam,
Science Park 904, 1098 XH Amsterdam, The Netherlands}\\
\mbox{$^{4}$ Russian Quantum Center, Novaya street 100, Skolkovo, Moscow region 143025, Russia}\\
\mbox{$^{5}$Wuhan Institute of Physics and Mathematics, Chinese Academy of Sciences, Wuhan 430071, China}} 

\date{\today}
\begin{abstract}
We consider the many-body localization-delocalization transition for strongly interacting one-dimensional disordered bosons and construct the full picture of finite temperature behavior of this system. This picture shows two insulator-fluid transitions at any finite temperature when varying the interaction strength. At weak interactions an increase in the interaction strength leads to insulator$\rightarrow$fluid transition, and for large interactions one has a reentrance to the insulator regime.
\end{abstract}

\maketitle
In spite of intensive studies during several decades, Anderson localization of quantum particles in disorder \cite{Anderson1958} remains one of the most active directions of research in condensed matter physics \cite{Abrahams}. A subtle question is how interactions between the particles affect the localization. It was first raised for electrons in solids \cite{Fleishman1980,BAA2006} and is now becoming crucial for ultracold neutral atoms in random potentials \cite{Modugno1,Modugno2,AAS2010,MAS2014}. After the first experiments on the observation of Anderson localization in expanding dilute quasi-one-dimensional (1D) clouds of bosonic atoms \cite{Billy2008,Roati2008}, the research on ultracold quantum gases in disorder rapidly grows.

Interacting quantum particles can undergo many-body localization-delocalization transition (MBLDT) - the transition from insulator to fluid state \cite{BAA2006}. MBLDT in a system of  disordered  weakly interacting 1D bosons has been discussed both at zero \cite{Altman2008,Falco2009,Lugan} and finite temperatures \cite{AAS2010}. In the latter case MBLDT manifests itself as a non-conventional insulator-normal fluid phase transition, the transport properties being singular at the transition point: in the fluid phase the mass transport is possible, whereas in the insulator phase it is completely blocked although the temperature $T$ is finite. The fluid-insulator transition for strong interactions at $T=0$ has been discussed by Giamarchi and Schulz \cite{Giamarchi1988}, who predicted that for the Luttinger liquid parameter $K<3/2$ even an arbitrary weak short-range disorder leads to localization.

Up to now the finite temperature behavior of 1D disordered bosons was well understood only for weak interactions. In this Letter we extend this understanding to the general case of strong and moderate interactions.  {\it At any finite temperature}, we show the presence of two insulator-fluid transitions: the insulator$\rightarrow$fluid transition when the interaction strength increases from zero to a certain critical value, and then a reentrance to the insulator phase at sufficiently strong interactions.

The physical picture can be interpreted as follows. Localization of all single-particle quantum eigenstates in 1D by an arbitrary weak disorder \cite{Gert,Mott} implies the insulating phase in the absence 
of the interaction between the bosons. In Ref. \cite{AAS2010} it was demonstrated that arbitrary weak interactions are unable to destroy the insulator: the boson density gets fragmented into lakes with irrelevant tunneling between them. The tunneling becomes relevant and drives the system into a fluid state at a critical interaction strength determined by the disorder. 

On the other hand, it is well known that in the absence of disorder bosons with an infinitely strong repulsion are equivalent to free fermions \cite{Girardeau1960} and, hence, they will be localized by an arbitrary weak disorder. Accordingly, there should be a second critical interaction strength above which one should expect an insulating state. In order to describe this transition using the 1D boson-fermion duality one should determine an effective interaction between the fermions when the boson-boson interaction is strong but finite. It is also important to account for the renormalization of the disorder by the fermion-fermion interaction even if the latter is fairly weak (see below).

The system of $N$ one-dimensional bosonic atoms repulsively interacting with each other via a short-range potential is well described by the Lieb-Liniger model \cite{LL1963}. The Hamiltonian reads:
\begin{equation}\label{HL}
 H_{B}=-\frac{\hbar^2}{2m}\sum_{j=1}^{N}\partial_{x_j}^2+g\sum_{1\leq j<k\leq N}\delta(x_j-x_k),
\end{equation}
where $g>0$ is the coupling constant, and $m$ the atom mass. The interaction strength is characterized by the dimensionless parameter 
\begin{equation}\label{gamma}
 \gamma=\frac{mg}{\hbar^2n},
\end{equation}
with $n=N/L$ being the average density of bosons, and $L$ the system length. In the regime of strong interactions we have $\gamma\gg 1$, and in the Tonks-Girardeau limit of an infinitly strong repulsion ($\gamma\rightarrow\infty$) the system maps onto the 1D gas of free fermions \cite{Girardeau1960}. In the general case of an arbitrary interaction strength, the Lieb-Liniger model (\ref{HL}) maps onto spinless (spin-polarized) fermions interacting with each other via an odd-wave momentum-dependent attractive interaction 
\cite{Cheon1999,Grosse2004,Sen,Cazalilla2011}. More precisely, the fermions are governed by the Hamiltonian 
\begin{equation}   \label{HF}
H_F=-\frac{\hbar^2}{2m}\sum_{j=1}^{N}\partial_{x_j}^2+V_F,
\end{equation} 
with the interaction operator \cite{Grosse2004}
\begin{equation}\label{VF}
 V_F=\frac{\hbar^4}{m^2g}\sum_{1\leq j<k\leq N}(\partial_{x_j}-\partial_{x_k})\delta(x_j-x_k)(\partial_{x_j}-\partial_{x_k}),
\end{equation}
and the eigenfunctions of $H_F$ (\ref{HF}) coincide with the bosonic eigenfunctions of $H_{B}$ (\ref{HL}) when the coordinates $x_j$ are ordered: $x_j<x_{j+1}$. Strongly repulsive bosons map onto weakly attractive spinless fermions with the Fermi momentum $k_F=\pi n$ and Fermi energy $E_F=\hbar^2k_F^2/2m$. In the degenerate regime the fermionic dimensionless coupling constant can be estimated as the ratio of a typical interaction energy (\ref{VF}) per particle to the Fermi energy:
\begin{equation}\label{dual}
\lambda\sim-\frac{\hbar^4k_F^3/m^2g}{E_F}\sim -\frac{1}{\gamma}\ll 1.
\end{equation}
Having in mind the comparison between strongly and weakly interacting bosons we use below the temperature of quantum degeneracy 
$$T_d=\frac{\pi^2\hbar^2 n^2}{2m},$$ 
which coincides with $E_F$ of effective fermions in the case of strongly interacting bosons.

Without loss of generality, we can represent the static disorder by a Gaussian random potential $U(x)$ with zero mean, variance $U_0^2$, and correlation length $\sigma$: 
$$\langle U(x)\rangle=0;\,\,\,\langle U(x)U(x')\rangle=U_0^2f(|x-x'|/\sigma),$$
where $f(z)\rightarrow 1$ for $z\rightarrow 0$, $f(z)\rightarrow 0$ for $z\rightarrow\infty$,
and the symbol $\langle...\rangle$ denotes averaging over the realizations of the disorder. 
Below we assume that $U_0\ll\hbar^2/m\sigma^2$. The energy scale brought by the disorder is \cite{IML,HL,ZL}
$$\varepsilon_\ast=\left(\frac{m\sigma^2 U_0^4}{\hbar^2}\right)^{1/3}\ll U_0.$$ 
Derivation of this equation, which was made in \cite{AAS2010}, is described in the Supplemental Material \cite{Suppl} in more detail.

For a weak disorder, $\varepsilon_\ast\ll T_d$, the localization length at characteristic energy $\varepsilon\sim T_d$, given by \cite{IML,HL,ZL}
\begin{equation}       \label{zeta}
\zeta(\varepsilon)\sim\frac{\hbar\varepsilon}{m^{1/2}\varepsilon_ \ast^{3/2}},
\end{equation}
greatly exceeds the mean interparticle distance $n^{-1}$. This ensures the presence of a small parameter:
\begin{equation}\label{D}
 \mathcal{D}=\left(\frac{\varepsilon_\ast}{T_d}\right)^{3/2}\ll 1.
\end{equation}

In the limit of $\gamma\rightarrow\infty$ the fermions are free \cite{Girardeau1960}, and an arbitrary weak disorder localizes them irrespective of their energy, leading to an insulator state at any temperature. A weak interaction between the fermions (and thus a strong but finite interaction between the bosons) causes many-body delocalization, i.e. destroys the insulator at the critical temperature (see Eq.(42) in \cite{BAA2006} and Supplemental Material \cite{Suppl})
\begin{equation}\label{Tc}
 T_c=\frac{C\delta_\zeta}{|\lambda\ln|\lambda||}.
\end{equation}
Here $C\sim 1$ is a model-dependent numerical constant, and $\delta_\zeta=\pi \hbar^2 k_F/m\zeta$ is the single-particle level spacing on the scale of the localization length $\zeta$. For a weak disorder in the 1D case $\zeta\approx \hbar k_F\tau/m$, with $\tau$ being the transport time. Thus we have:
\begin{equation}\label{delta}
 \delta_\zeta\approx\frac{\pi\hbar}{\tau}.
\end{equation}

It is known that the interaction between 1D fermions renormalizes the disorder (see Refs. \cite{DisRenorm,Matveev}):
\begin{equation}\label{tau}
 \frac{1}{\tau}=\frac{1}{\tau_0}\left(\frac{T_d}{T}\right)^{2\lambda};\,\,\,\,|\lambda|\ll 1,\,\,T\ll T_d.
\end{equation}
Here $1/\tau_0$ is the particle-impurity scattering rate. In the Born approximation (see Supplemental
Material \cite{Suppl})
\begin{equation}   \label{tau0}
\frac{1}{\tau_0}=\frac{2m\sigma U_0^2}{\pi n\hbar^3}=\frac{\varepsilon_\ast}{\hbar}\left(\frac{2\varepsilon_\ast}{T_d}\right)^{1/2}.
\end{equation}

The dimensionless fermionic coupling constant is defined \cite{Matveev} as:
\begin{equation}\label{alpha}
 \lambda=\frac{\big[\tilde{V}(0)-\tilde{V}(2k_F)\big]m}{2\pi^2\hbar^2n},
\end{equation}
where $\tilde{V}(q)$ is the Fourier transform of the interaction potential $V_F$ (\ref{VF}) for the transferred momentum $q$. In the secondly quantized form the potential (\ref{VF}) writes:
\begin{eqnarray}
V_F^{(2)}&=&\frac{1}{2L}\sum_{k_1,k_2,q}\tilde V(q,k_1,k_2)    \nonumber \\
&&a^\dagger(k_1+q)a^\dagger(k_2-q)a(k_2)a(k_1),\label{VFkop}
\end{eqnarray}
where the operator $a(k)$ annihilates a fermion with momentum $k$, and 
\begin{equation}  \label{VFk}
\tilde V(q,k_1,k_2)=-\frac{\hbar^4}{m^2g}(k_1-k_2+2q)(k_1-k_2).
\end{equation}
For $q=2k_F$ the phase-space constraint fixes $k_1=-k_2=-k_F$, and for $q=0$ we have $(k_1-k_2)=\pm 2k_F$. This leads to \cite{Vq}
\begin{equation}   \label{VV}
 \tilde{V}(2k_F)-\tilde{V}(0)=\frac{8\hbar^4k_F^2}{m^2g}=\frac{8\pi^2\hbar^2 n}{m\gamma}.
\end{equation}
Substitution of Eq.(\ref{VV}) into Eq.(\ref{alpha}) yields the fermionic coupling constant 
\begin{equation}   \label{alphagamma}
 \lambda=-\frac{4}{\gamma}<0.
\end{equation}

Thus, according to Eqs.(\ref{tau}) and (\ref{delta}) the renormalized impurity backscattering rate and the effective level spacing decrease with temperature due to the interfermion interaction (\ref{VF}). Using Eqs. (\ref{D}), (\ref{tau0}), and (\ref{alphagamma}) we find:
\begin{equation}\label{tauresult}
 \frac{1}{\tau}=\sqrt{2}\mathcal{D}\frac{T_d}{\hbar}\Big(\frac{T}{T_d}\Big)^{8/\gamma}.
\end{equation}
Inserting Eq.(\ref{tauresult}) into Eq.(\ref{delta}) and using Eq.(\ref{Tc}) we obtain a critical temperature in the quantum degenerate regime:
\begin{equation}\label{Tcdeg}
 T_c\sim T_d\Big(\frac{\mathcal{D}\gamma}{\ln\gamma}\Big)^\frac{\gamma}{\gamma-8};\,\,\,\frac{\gamma}{\ln\gamma}\ll\frac{1}{\mathcal{D}}.
\end{equation}
Equation (\ref{Tcdeg}) indicates that $T_c$ vanishes at $\gamma=\gamma_0\approx8$. Note that the weak disorder renormalization group approach of Giamarchi-Schulz \cite{Giamarchi1988} yields the critical Luttinger parameter $K=3/2$ corresponding to $\gamma_0=7.9$, so that at $T=0$ an infinitesimally small disorder leads to localization (insulator state) for $K<3/2$ ($\gamma>7.9$). In this sense, our equation (\ref{Tcdeg}) is in agreement with the zero temperature result of Ref. \cite{Giamarchi1988}. 

In the derivation of Eq.(\ref{Tcdeg}) we took into account the renormalization of the disorder due to interaction of the effective spinless fermions with Friedel oscillations and neglected the renormalization of the interaction by the disorder. It is known \cite{Giamarchi1988} that this approximation works as long as $(\gamma-\gamma_0)\gg\mathcal{D}^{1/2}$. Close to the zero-temperature transition this condition is violated and one has to use the coupled renormalization group equations for the interaction and disorder. For $\gamma<\gamma_0$ this leads to the Berezinskii-Kosterlitz-Thouless (BKT) criticality with temperature (see the Supplemental Material \cite{Suppl})
\begin{equation}  \label{TBKT}
T_c\sim T_d\exp[-16\pi/\sqrt{c_1^2\mathcal{D}-c_2^2(\gamma-\gamma_0)^2}], 
\end{equation}
where $c_1\approx 8.5$ and $c_2\approx0.93$. As soon as $\gamma$ becomes larger than $\gamma_0$ and $(\gamma-\gamma_0)$ exceeds $\mathcal{D}^{1/2}$, the low-temperature behavior obeys Eq.(\ref{Tcdeg}).

The derivation of Eq.(\ref{TBKT}) is similar to that for the localization length scale $L_{loc}$ in the case of spin-1/2 fermions in Ref. \cite{Giamarchi1988}. The relation $T_c\sim \hbar v_F/L_{loc}$, where $v_F$ is the Fermi velocity, is the same as Eq.(\ref{TBKT}), except for the coefficients $c_1$ and $c_2$ because the critical Luttinger parameter for spin-1/2 fermions is equal to $3$. However, this $T_c$ was treated as a temperature of the crossover from low to high resistivity. We claim that $T_c$ is the insulator-fluid phase transition temperature.  

In the limit of extremely strong coupling, $\gamma\gg 8$, the power of $\gamma$ in Eq.(\ref{Tcdeg}) becomes equal to 1, which gives 
\begin{equation}   \label{Igor}
T_c\sim T_d\frac{\mathcal{D}\gamma}{\ln\gamma}.
\end{equation} 
The substitution of $\gamma\sim 1/|\lambda|$ from Eq.(\ref{dual}) into Eq.(\ref{Igor}) transforms it to the result of Ref.\cite{BAA2006} for $T_c$ of weakly interacting spinless fermions. Note that in this regime the critical temperature does not depend on the sign of the interaction between the fermions.

For a non-degenerate gas where $T_d\ll T$, the transition  temperature can be found by using the single-particle picture of Ref. \cite{AAS2010} after mapping strongly interacting bosons onto weakly interacting fermions. We then have:
\begin{equation} \label{deltacl}
 \delta_\zeta(T_c)\sim n\tilde{V}(T_c).
\end{equation}
The level spacing is $\delta_\zeta(\varepsilon)\sim 1/\nu(\varepsilon)\zeta(\varepsilon)$, where $\nu(\varepsilon)\sim\sqrt{m/\hbar^2\varepsilon}$ is the density of states, and the localization length is given by Eq.(\ref{zeta}). According to Eq.(\ref{VFk}) we have $\tilde{V}(\varepsilon)\sim\hbar^2\varepsilon/mg$, so that $n\tilde{V}(T)/T\sim1/\gamma\ll1$. This leads to
\begin{equation}\label{Tcdil}
 T_c\sim T_d(\mathcal{D}\gamma)^{2/3}\sim\varepsilon_\ast\gamma^{2/3};\,\,\,\,\,T\gg T_d.
\end{equation}
The value of $\gamma$ corresponding to a crossover from the classical to quantum degenerate regime is obtained from equation (\ref{Tcdil}) setting $T_c\sim T_d$. This yields  $\gamma\sim1/\mathcal{D}\gg1$.

It should be noted that at temperatures of the order of $T_c$ (\ref{Tcdil}) the fermions remain weakly interacting. The scattering phase shift $\delta_F(k)$ in the two-body problem with the interaction potential (\ref{VF}) is 
\begin{equation}
 \delta_F(k)=-\arctan\Big(\frac{\hbar^2 k}{2mg}\Big)=-\arctan\Big(\frac{k}{2\gamma n}\Big).                                                                                                                                                                                                                                                                                                                                                                                                                                                         \end{equation}
For $T<T_d$ the fermions are degenerate and $k\sim k_F\sim n$. The phase shift $\delta_F(k)\sim-\arctan(\gamma^{-1})$ is thus small as long as $\gamma\gg 1$, i.e. the fermions interact weakly. In the non-degenerate case,  $T\gg T_d$, the momentum $k$ is of the order of the thermal momentum $\sqrt{mT/\hbar^2}$ and at very high temperatures the fermions interact strongly, $\delta_F\sim 1$. However, at $T\lesssim T_c$ (\ref{Tcdil}) the phase shift is still small, $\delta_F\leq\mathcal{D}^{1/3}/\gamma^{2/3}\ll1$, and the MBLDT single-particle picture leading to equation (\ref{Tcdil}) is valid.

In Fig.1 we show the phase diagram in terms of the amplitude of the disorder ($\mathcal{D}$) and temperature at a fixed interaction strength ($\gamma$).  As expected, the critical disorder for the fluid-insulator transition vanishes as $T\rightarrow 0$ if $\gamma>8$, whereas for $\gamma<8$ it remains finite.

\begin{figure}
\includegraphics[width=9cm]{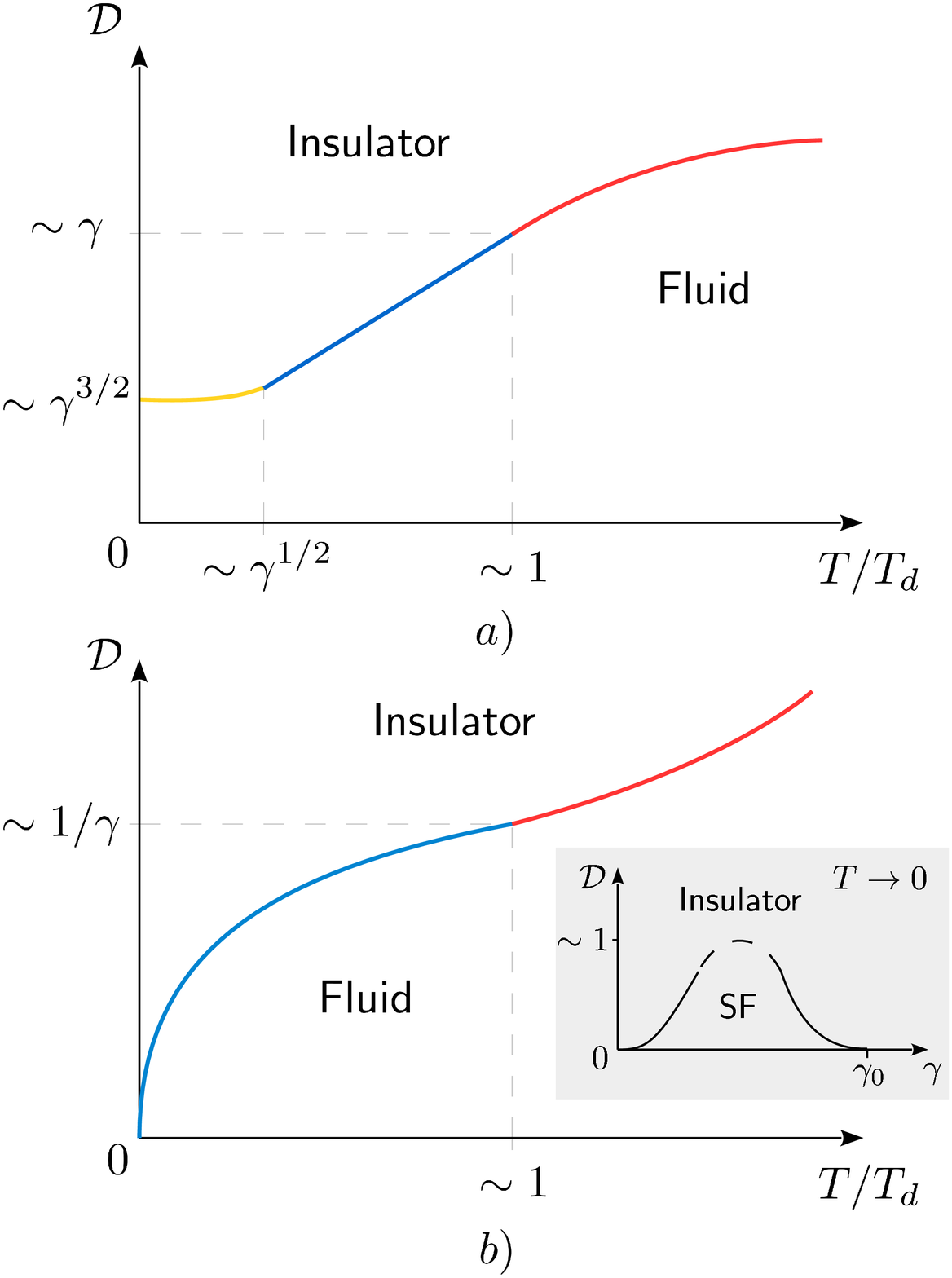}
\caption{Phase diagram in terms of the disorder strength $\mathcal{D}$ and temperature at a fixed dimensionless coupling constant $\gamma$. In a) $\gamma\ll 1$. The yellow part of the curve corresponds to Eq.(\ref{1}), the blue part to Eq.(\ref{2}), and the red part to Eq.(\ref{3}). In b) $\gamma>\gamma_0\approx 8$, and the blue part of the curve corresponds to Eq.(\ref{Tcdeg}), whereas the red part to Eq.(\ref{Tcdil}). Inset: dependence of the critical disorder on $\gamma$ at
$T\to0$, described by Eq.(\ref{1}) for $\gamma\ll1$ and by Eq.(\ref{4}) for $\gamma$ close to $\gamma_0\approx8$.
The dashed part of the curve shows the expected behavior for $\mathcal{D}\sim 1$.}
\label{Fig1}
\end{figure}

The phase diagram in terms of dimensionless temperature $T/T_d$ and interaction strength $\gamma$ at a fixed weak disorder ($\mathcal{D}\ll 1$) is displayed in Fig.2. It combines the results of the present paper for strong coupling of the bosons, $\gamma\gg 1$, with the weak coupling ($\gamma\ll 1$) results of Ref. \cite{AAS2010}. We clearly see that in a weak disorder ($\mathcal{D}\ll 1$) at any finite temperature one has two insulator-fluid transitions as $\gamma$ increases from small to large values. The first insulator to fluid transition occurs when the interaction between the bosons is weak ($\gamma\ll 1$). The results of Ref. \cite{AAS2010} for this case can be written in terms of the critical coupling $\gamma_{c1}$ of the transition as:
\begin{eqnarray}
&\gamma_{c1}&\sim\mathcal{D}^{2/3};\,\,\,\,\,T\ll \mathcal{D}^{1/3}T_d, \label{1} \\
&\gamma_{c1}&\sim \mathcal{D}T_d/T;\,\,\,\,\,\mathcal{D}^{1/3}T_d\ll T\ll T_d, \label{2} \\
&\gamma_{c1}&\sim \mathcal{D}(T_d/T)^{1/2};\,\,\,\,\,T\gg T_d. \label{3}
\end{eqnarray}
The reentrance to the insulator takes place at strong interactions ($\gamma\gg 1$): 
\begin{eqnarray}
&\gamma_{c2}&=\gamma_0-c\sqrt{\mathcal{D}};\,\,\,\,\,T\rightarrow 0, \label{4} \\
&\gamma_{c2}&\sim \mathcal{D}^{-1}(T/T_d)^{3/2};\,\,\,\,\,T\gg T_d. \label{5}
\end{eqnarray}
For finite temperatures $T\ll T_d$ the critical coupling is determined by a transcendental equation following from Eq.(\ref{Tcdeg}). The $T\rightarrow 0$ asymptotics corresponds to the 1+1 BKT quantum transition only slightly modified by the weak disorder. In the limit of $\mathcal{D}\rightarrow 0$ equation (\ref{4}) thus  reproduces the result of Giamarchi and Schulz \cite{Giamarchi1988}. 

\begin{figure}
\includegraphics[width=9cm]{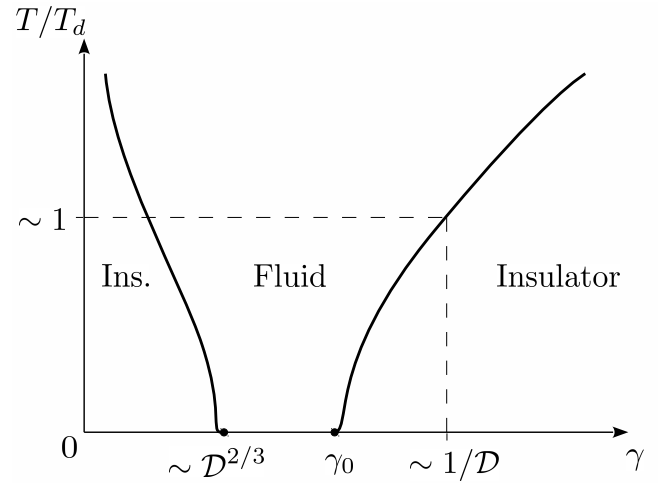}
\caption{Fluid-insulator transition of finite temperature repulsive bosons in a weak disorder, $\mathcal{D}\ll1$ (see text).}
\label{Fig2}
\end{figure}

The difference $\gamma_{c2}-\gamma_{c1}$ between the two critical interaction constants should decrease as the disorder ($\mathcal{D}$) increases. The two insulating regimes at $T=0$ are probably merging ($\gamma_{c1}=\gamma_{c2}$) at a certain critical value of the disorder, $\mathcal{D}_c\sim 1$. 

It is feasible to verify experimentally the full picture of the finite temperature behavior of 1D disordered bosons constructed in the present paper. A suitable candidate would be the gas of $^7$Li atoms where the coupling constant $g$ can be varied by Feshbach resonance from very small to very large values \cite{Hulet1}, and the 1D regime has already been achieved \cite{Salomon,Hulet2}. The regime of strong interactions can be reached, in particular, by using a confinement-induced resonance as in the cesium experiments \cite{Haller}, and the disorder can be introduced by using optical speckles like in the first experiments on the observation of Anderson localization \cite{Billy2008}. The insulator-fluid transition can be identified in expansion experiments (see the discussion in Ref. \cite{AAS2010}), or by analyzing the momentum distribution like in recent experiments for bosons in 1D quasiperiodic potentials \cite{Modugno1,Modugno2}.  

{\it Note Added}\\
After the present work has been finished, we learned about Ref. \cite{BA} where the two insulator-fluid transitions were observed in the experiment with fermions in the one-dimensional quasiperiodic potential.

We are grateful to M.B. Zvonarev and A.Iu. Gudyma for fruitful discussions and acknowledge support from IFRAF and from the Dutch Foundation FOM. The research leading to these results has received funding from the European Research Council under European Community's Seventh Framework Programme (FR7/2007-2013 Grant Agreement no.341197).

\pagebreak
\widetext
\begin{center}
\textbf{\large Supplemental Material: Finite-Temperature Fluid-Insulator Transition of Strongly Interacting 1D Disordered Bosons}
\end{center}
\setcounter{equation}{0}
\setcounter{figure}{0}
\setcounter{table}{0}
\setcounter{page}{1}
\makeatletter
\renewcommand{\theequation}{S\arabic{equation}}
\renewcommand{\thefigure}{S\arabic{figure}}
\renewcommand{\bibnumfmt}[1]{[S#1]}
\renewcommand{\citenumfont}[1]{S#1}

We detail here five points of the paper. We begin with the description of the disorder potential and the length and energy scales of the 1D localized states in the absence of interaction, then we explain the fermion/boson duality in 1D in the presence of an external potential. Furthermore we characterize the finite-temperature fluid-insulator transition of weakly interacting spinless fermions in the degenerate and dilute regimes. In the fourth part we detail the renormalization of the disorder strength due to weak interfermion interaction. Finally we point out the relation between our results and the Giamarchi/Schulz renormalization group formulation of the weakly disordered Luttinger liquid.

{\it Disorder potential length and energy scales.} We consider a Gaussian random potential $U(x)$ with the correlation length $\sigma$ and the amplitude $U_0$ (see the main text). The energy-dependence of the localization length is known \cite{S_IML,S_HL,S_ZL}:
\begin{equation}\label{Szeta}
 \zeta(\varepsilon)\approx\hbar^2\varepsilon/m\sigma U_0^2,
\end{equation}
under the condition
\begin{equation}\label{cond}
\zeta\gg\sigma;
\end{equation}
so that the extent of the wave-function covers a distance over which the random potential varies many times. 
Let us estimate the energy of a low-lying bound state in the random potential. The kinetic energy of the state is 
\begin{equation}
 K\sim\hbar^2/2m\zeta^2,
\end{equation}
and the potential energy is computed as follows. Consider the segment of the length $\zeta$, which is made of fragments of size $\sigma$, where the value of the potential is $\sim U_0$. The number of such fragments is $\zeta/\sigma$. A wavefunction with the extent $\zeta$ gives the potential energy contribution of a typical fragment $\sim U_0\sigma/\zeta$. In the case of the Gaussian fluctuations the typical potential energy of the state is the energy contribution of each fragment multiplied by the square root of the number of fragments $\sqrt{\zeta/\sigma}$, that is
\begin{equation}
 U\sim-U_0\sigma/\zeta\times\sqrt{\zeta/\sigma}=-U_0\sqrt{\sigma/\zeta}.
\end{equation}
Minimizing the total energy $E=K+U$ with respect to $\zeta$ and taking into account Eq.(\ref{Szeta}) one obtains the length and energy scales of the low-energy bound state
\begin{equation} 
 \zeta_\ast=(\hbar^4/m^2\sigma U_0^2)^{1/3},\quad\varepsilon_\ast=(m\sigma^2U_0^4/\hbar^2)^{1/3}.
\end{equation}
Inserting $\zeta_\ast$ into Eq.(\ref{cond}) directly leads to the condition stated in the main text:
\begin{equation}\label{cond2}
 U_0\ll\hbar^2/m\sigma^2,
\end{equation}
and one then clearly sees that $\varepsilon_\ast\ll U_0$.

{\it Boson/Fermion correspondence in 1D}. 
Here we extend the arguments of Ref.\cite{S_Grosse2004} to the case of an external (random) potential. We demonstrate that although the model is not exactly solvable the 1D fermion-boson duality remains valid.

Consider the Schr\"odinger equation for the $N$-fermion wavefunction $\chi(x_1,\dots,x_N)$:
\begin{equation}\label{SE}
 (H_1+V_F)\chi(x_1,\dots,x_N)=E\chi(x_1,\dots,x_N).
\end{equation}
Here $H_1$ is the single-particle part of the Hamiltonian, which differs from Eq.(3) of the main text by the presence of the external (disorder) potential $U(x)$:
\begin{equation}\label{H1}
 H_1=\sum_{j=1}^N\Big[-\frac{\hbar^2}{2m}\partial_{x_j}^2+U(x_j)\Big].
\end{equation}
The two-body interaction part of the Hamiltonian, $V_F$, is determined by Eq.(4) of the main text, which can be rewritten as 
\begin{equation}\label{SVF}
 V_F=\frac{4\hbar^4}{m^2g}\sum_{1\leq j<k\leq N}\partial_{jk}\delta(x_{jk})\partial_{jk},
\end{equation}
where $x_{jk}=x_j-x_k$ and $\partial_{jk}=(\partial_{x_j}-\partial_{x_k})/2$.

For non-coinciding coordinates Eq.(\ref{SE}) reduces to
\begin{equation}\label{SE2}
 H_1\chi(x_1,\dots,x_N)=E\chi(x_1,\dots,x_N).
\end{equation}
This equation should be supplemented by the boundary conditions, which determine the behavior of the wavefunction in the limits $x_j\to x_k\pm0$. In order to find these conditions one integrates Eq.(\ref{SE}) over an infinitesimally small interval $-\tilde x<x_{jk}<x$; $\tilde x,x\to+0$ of $x_{jk}$. Using  Eqs.(\ref{H1}) and (\ref{SVF}) and taking into account the identity $\partial_{x_j}^2+\partial_{x_k}^2=2\partial_{jk}^2+\frac{1}{2}\partial_{X_{jk}}^2$ (here $X_{jk}=(x_j+x_k)/2$) we then obtain:
\begin{equation}
 \int_{-\tilde x}^{x} dx_{jk}[-\frac{\hbar^2}{m}\partial_{jk}^2\!+\!\frac{4\hbar^4}{m^2g}\partial_{jk}\delta(x_{jk})\partial_{jk}]\chi(x_{jk},\dots)+\textnormal{Regular}=0, 
\end{equation}
where ``Regular'' includes terms that vanish in the limit $\tilde x,x\to0$.
Performing the integration in the first term we obtain the condition which does not contain the external (random) potential:
\begin{equation} \label{int}
-\frac{\hbar^2}{m}\partial_x[\chi(x,\dots)-\chi(-\tilde x,\dots)]+\frac{4\hbar^4}{m^2g}[\delta(x)\partial_{x}\chi(x,\dots)-\delta(-\tilde x)\partial_{x}\chi(-\tilde x,\dots)]=0. 
\end{equation}
In the limit $\tilde x,x\to+0$ equation (\ref{int}) can be written as
\begin{equation}
 \partial_{jk}\chi(x_{jk}=+0,\dots)=\partial_{jk}\chi(x_{jk}=-0,\dots).
\end{equation}

Integrating Eq.(\ref{int}) over $x$ in an infinitesimally small interval $-\tilde x<x<\tilde x$ and taking the limit $\tilde x\to +0$ we have:
\begin{equation}\label{BC2inter}
 \chi(x_{jk}=+0,\dots)-\chi(x_{jk}=-0,\dots)=\frac{4\hbar^2}{mg}\partial_{jk}\chi(x_{jk},\dots)|_{x_{jk}=+0}.
\end{equation}
The fermionic wavefunction is antisymmetric with respect to the transmutation of $x_j$ and $x_k$. Accordingly, $\chi(x_{jk}=+0,\dots)=-\chi(x_{jk}=-0,\dots)$, and in terms of $x_j,x_k$ equation (\ref{BC2inter}) takes the form:
\begin{equation}\label{BC2}
 \Big(\partial_{x_j}-\partial_{x_k}-\frac{mg}{\hbar^2}\Big)\chi(\dots,x_j,x_k,\dots)|_{x_j=x_k+0}=0.
\end{equation}

Note that Eqs.(\ref{SE2}) and (\ref{BC2}) could equally be obtained for the Lieb-Liniger Hamiltonian of Eq.(1) of the main text, completed by the external potential:
\begin{equation}\label{HB}
 H_B=H_1+g\sum_{j<k}\delta(x_j-x_k),
\end{equation}
which acts on the N-body bosonic wavefunctions. As a consequence the many-body spectra of the models (\ref{SE}) and (\ref{HB}) are identical and the eigenfunctions are equal in magnitude and only differ by a sign upon exchange of coordinates.

{\it Conductor-insulator transition of disordered weakly interacting 1D spinless fermions with gradient-dependent interaction}.
In the case where all single-particle eigenstates are localized (as in 1D), there is a critical temperature $T_c$ above which weakly interacting spinless fermions turn to a conducting state. Here we characterize the transition both in the quantum degenerate regime \cite{S_BAA2006} and in the high-temperature dilute regime \cite{S_AAS2010}. In the basis of the localized single-particle states $\psi_\alpha(x)$ with energies $\varepsilon_\alpha$ the many-body Hamiltonian $H_1+V_F$, with $V_F$ given by Eq.(\ref{SVF}), writes
\begin{equation}\label{H}
 H_F=\sum_\alpha\varepsilon_\alpha a_\alpha^\dagger a_\alpha+\frac{1}{2}\sum_{\alpha\beta\gamma\delta}V_{\alpha\beta}^{\gamma\delta}a_\gamma^\dagger a_\delta^\dagger a_\beta a_\alpha,
\end{equation}
where $a_\alpha$ are the field operators and the two-body interaction matrix elements $V_{\alpha\beta}^{\gamma\delta}$ are given by:
\begin{equation}\label{ME}
 V_{\alpha\beta}^{\gamma\delta}\!=\!-\frac{4
 \hbar^4}{m^2g}\int dx\psi^*_\gamma(x)[\partial_x\psi^*_\delta(x)][\partial_x\psi_\beta(x)]\psi_\alpha(x).
\end{equation}
The single-particle states take the form
\begin{equation}
 \psi_\alpha(x)=A_\alpha(x)e^{-|x-x_\alpha|/\zeta_\alpha},
\end{equation}
where $A_\alpha(x)$ is an oscillating function with wave number
$$k_\alpha=\sqrt{2m\varepsilon_\alpha}/\hbar.$$
Therefore the matrix elements (\ref{ME}) evaluate to
\begin{equation}
\nonumber V_{\alpha\beta}^{\gamma\delta}\approx\frac{4\hbar^4k_\beta k_\delta}{m^2g}\int dx A_\gamma(x)A_\delta(x)A_\beta(x)A_\alpha(x)e^{-(|x-x_\gamma|/\zeta_\gamma+|x-x_\delta|/\zeta_\delta+|x-x_\beta|/\zeta_\beta+|x-x_\alpha|/\zeta_\alpha)},
\end{equation}
Taking into account the condition that the levels $\alpha$,$\gamma$ and $\beta$,$\delta$ are neighbors in energy \cite{S_BAA2006,S_AAS2010}, we obtain
\begin{equation}
V_{\alpha\beta}^{\gamma\delta}\approx\frac{4\hbar^4k_\beta^2}{m^2g\max(\zeta_\alpha,\zeta_\beta)}=\frac{8\hbar^2\varepsilon_\beta}{mg\max(\zeta_\alpha,\zeta_\beta)}.
\end{equation}

In the quantum degenerate regime, $T\ll T_d=\pi^2\hbar^2n^2/2m$ ($n$ is the average density of particles), the two-body processes responsible for the fluid-insulator transition occur at the energy scale $T_d$. The relevant two-body matrix elements are
\begin{equation}\label{ME2}
 V_{\alpha\beta}^{\gamma\delta}\sim\lambda\delta_\zeta,
\end{equation}
where the effective interfermion coupling parameter $\lambda$ is of the order of the inverse of the Lieb-Liniger parameter: 
\begin{equation}\label{Sdual}
 \lambda\sim\frac{\hbar^2n}{mg}\equiv\frac{1}{\gamma}\ll1,
\end{equation}
and the spacing between the single-particle energy levels in a segment of length $\zeta$ writes
\begin{equation}
\delta_\zeta=\frac{1}{\nu\zeta},
\end{equation}
with $\nu=m/\pi^2\hbar^2 n$ being the density of states at the Fermi energy.
The many-body Hamiltonian (\ref{H}) with the two-body matrix elements (\ref{ME2}) leads to the critical temperature in the quantum degenerate regime \cite{S_BAA2006}:
\begin{equation}
 T_c=\frac{C\delta_\zeta}{|\lambda\ln|\lambda||},
\end{equation}
where $C$ is a model-dependent numerical constant of the order one. Taking into account the duality (\ref{Sdual}) and the level spacing renormalization due to interparticle interaction (described in the next part), we obtain Eq.(18) of the main text.

In the dilute high-temperature case $T\gg T_d$, the average occupation of the single-particle energy levels is small and given by the Boltzmann distribution. In that case the temperature $T$ gives the relevant energy scale of the states participating in the interaction processes. The critical temperature in this regime is given by \cite{S_AAS2010}
\begin{equation}\label{STcdil}
 \delta_\zeta(T_c)\sim n\tilde{V}(T_c),
\end{equation}
with the one-particle level spacing at energy $\varepsilon$:
\begin{equation}\label{delta}
 \delta_\zeta(\varepsilon)=\frac{1}{\nu(\varepsilon)\zeta(\varepsilon)}.
\end{equation}
The energy-dependence of the density of states is
\begin{equation}\label{nu}
 \nu(\varepsilon)=\sqrt{m/2\pi^2\hbar^2\varepsilon},
\end{equation}
and the localization length $\zeta(\varepsilon)$ is given by Eq.(\ref{Szeta}). In Eq.(\ref{STcdil}) the quantity $\tilde{V}(\varepsilon)$ is the Fourier transform of the two-body interaction potential at momentum $k=\sqrt{2m\varepsilon}/\hbar$:
\begin{equation}\label{Vtilde}
 \tilde{V}(\varepsilon)\approx\frac{4\hbar^2\varepsilon}{mg}.
\end{equation}
Inserting Eqs.(\ref{Szeta}), (\ref{nu}), (\ref{delta}), and (\ref{Vtilde}) into (\ref{STcdil}) we obtain Eq.(22) of the main text. 

{\it Renormalization of the impurity scattering amplitude.} In this section we describe the renormalization of the disorder strength by the interaction and derive the resulting Born approximation backscattering rate in the case of many impurities. 
The renormalization of a single impurity transmission amplitude $t$ due to weak interaction between spinless fermions was computed by Matveev et al. \cite{S_Matveev}. The result is a temperature dependence of the transmission coefficient $\mathcal{T}=|t|^2$:
\begin{equation}\label{renorm}
\mathcal{T}(T)=\frac{1}{1+(1/\mathcal{T}_0-1)(T_d/T)^{2\lambda}},
\end{equation}
here $\mathcal{T}_0$ is the impurity transmission coefficient in the absence of interactions. The coupling parameter is defined as
\begin{equation}\label{Slambda}
  \lambda=\frac{\tilde{V}(0)-\tilde{V}(2k_F)}{2\pi\hbar v_F}\ll1,
\end{equation}
with $\tilde{V}(k)$ being the Fourier transform of the interfermion potential, and $v_F=\pi\hbar n/m$ the Fermi velocity.

In quantum mechanics \cite{S_LLQM} a particle of energy $T_d$ has an amplitude of transmission through a potential of extent $\sigma$ and amplitude $U_0$ given by
\begin{equation}\label{renormTau}
 1/\mathcal{T}-1=m\sigma^2U_0^2/2\hbar^2T_d
\end{equation}
under the condition (\ref{cond2}).
On the other hand the backscattering rate caused by many impurities with concentration $c_{imp}$ is evaluated in the Born approximation ($T\tau\gg\hbar$) as
\begin{equation}
 1/\tau=2 c_{imp}\sigma^2 U_0^2/\hbar^2 v_F,
\end{equation}
and taking into account Eq.(\ref{renormTau}) we get
\begin{equation}\label{taucimp}
 1/\tau=\frac{2\pi\hbar nc_{imp}}{m}(1/\mathcal{T}-1).
\end{equation}
The prefactor in the right-hand side of Eq.(\ref{taucimp}) does not depend on the properties of the single impurity potential and therefore it is not affected by the renormalization (\ref{renorm}). As a consequence we can write the renormalized backscattering rate as
\begin{equation}
 1/\tau=(1/\tau_0)(T_d/T)^{2\lambda},
\end{equation}
where setting $c_{imp}=1/\sigma$ the bare backscattering time is 
\begin{equation}
\tau_0=\frac{\pi \hbar^3n}{2m\sigma U_0^2}.
\end{equation}

{\it Giamarchi and Schulz renormalization group for the disordered bosons.} We now turn to the Luttinger liquid formulation of the disordered 1D bosons and point out the relation between its predictions and our theory. The renormalization group approach of the disordered Luttinger liquid of bosons was developed by Giamarchi and Schulz \cite{S_Giamarchi1988} and it describes the zero-temperature algebraic-superfluid to Bose-glass transition at moderate interactions and small disorder.
The influence of disorder is taken into account by the backscattering field $\xi(x)$ \cite{S_Giamarchi}, with correlations
\begin{equation}
 \langle \xi(x)\xi^*(x')\rangle=\pi^2\hbar^2v^2n\tilde{\mathcal{D}}\delta(x-x'),
\end{equation}
where the symbol $\langle.\rangle$ represents the average over disorder realizations, $v=v_F/K$ is the velocity of the sound mode, $K$ is the Luttinger parameter, and $\tilde{\mathcal{D}}\ll1$ is the dimensionless parameter of the disorder. The relation between $\tilde{\mathcal{D}}$ and the parameter $\mathcal{D}$ that we use in the main text is
\begin{equation}
 \tilde{\mathcal{D}}=\frac{K^2}{\pi}\Big(\frac{\varepsilon_\ast}{2T_d}\Big)^{3/2}=\frac{K^2}{2^{3/2}\pi}\mathcal{D}.
\end{equation}

The scaling transformation of the disordered Luttinger liquid model leads to two coupled differential equations giving the flow of $K$ and $\tilde{\mathcal{D}}$ with respect to the scale parameter. The solution of the equations leads to the emergence in the localized phase of the length scale $L_{loc}$ which diverges at the glass-superfluid transition. There are two limiting cases where $L_{loc}$ takes on two different forms. 
When $K<3/2$ and $\tilde{\mathcal{D}}^{1/2}\ll 3/2-K$ this length scale writes
\begin{equation}
 L_{loc}\sim a(1/\tilde{\mathcal{D}})^{\frac{1}{3/2-K}},
\end{equation}
where $a$ is a short-distance cut-off that we set equal to $n^{-1}$.
Thus for infinitesimally small disorder the theory predicts the fluid-insulator transition at $K=3/2$.
At this point the parameter of the disorder strength is
\begin{equation}
 \tilde{\mathcal{D}}=\frac{9}{2^{7/2}\pi}\mathcal{D}\approx0.25\mathcal{D}.
\end{equation}
Having obtained the length scale $L_{loc}$ one can infer the critical temperature \cite{S_Comment2} 
\begin{equation}\label{Tloc}
 T_{c}\sim \hbar v_F/L_{loc}.
\end{equation}
It is interesting to note the correspondence between Eq.(\ref{Tloc}) and our strong coupling critical temperature in the degenerate regime given in the main text. Indeed given the development of the Luttinger parameter with respect to $1/\gamma=\hbar^2n/mg\ll1$:
\begin{equation}
K=1+4/\gamma+O(1/\gamma^2),
\end{equation}
we see that our case corresponds to the first-order strong coupling expansion of $K$ and our result extends Eq.(\ref{Tloc}) to the whole temperature regime $T<T_d=\pi^2\hbar^2n^2/2m$. The two results are in agreement in the regime  $\mathcal{D}^{1/2}\ll(\gamma-\gamma_0)/\gamma_0<1$ and the upper limit is equivalent to $T/T_d\ll\mathcal{D}$, where $\gamma_0\approx7.91(1)$ is the solution of the equation $K(\gamma)=3/2$ evaluated numerically from the Bethe Ansatz equations \cite{S_Lieb}.

On the other hand when $\tilde{\mathcal{D}}^{1/2}\gg K-3/2$ ($K>3/2$) the renormalization of the interaction is important and we must solve the coupled differential equations for $K$ and $\tilde{\mathcal{D}}$. This leads to the length scale
\begin{equation}
L_{loc}\sim a\exp\Big[\pi/\sqrt{9\tilde{\mathcal{D}}/8-(K-3/2)^2}\Big],
\end{equation}
and accounting for Eq.(\ref{Tloc}) and the dependence $K(\gamma)$ near $K=3/2$: $K(\gamma)\approx3/2+0.058(\gamma_0-\gamma)$, we obtain Eq.(19) of the main text.

\end{document}